\documentclass[journal,twocolumn]{IEEEtran}
\usepackage{graphicx,cite,calc,dsfont}
\usepackage[defs]{ams}
\usepackage{eqnarray,subeqn,xcolor}

\begin{document}

\title{One-Hop Throughput of Wireless Networks with Random Connections}

\author{ Seyed Pooya Shariatpanahi, Babak Hossein Khalaj, Kasra Alishahi, Hamed Shah-Mansouri 
 }

\maketitle
\thispagestyle{empty}
\pagestyle{empty}
\section*{Abstract}

\emph{\let\thefootnote\relax\footnotetext{  S. P. Shariatpanahi, B. H. Khalaj and H. Shah-Mansouri are with the Department of Electrical Engineering and Advanced Communication Research Institute (ACRI). K. Alishahi is with the Department of Mathematical Sciences, Sharif University of Technology, Azadi Ave., Tehran, Iran. Emails: pooya@ee.sharif.edu, khalaj@sharif.edu, alishahi@sharif.edu, hshahmansour@ee.sharif.ir.}
We consider one-hop communication in wireless networks with random connections. In the random connection model, the channel powers between different nodes are drawn from a common distribution in an i.i.d. manner. An scheme achieving the throughput scaling of order $n^{1/3-\delta}$, for any $\delta>0$, is proposed, where $n$ is the number of nodes. Such achievable throughput, along with the order $n^{1/3}$ upper bound derived by Cui et al., characterizes the throughput capacity of one-hop schemes for the class of connection models with finite mean and variance. \\ \\  \textbf{Keywords}: Wireless Networks, Random Connection Model, Achievable Throughput.}

\section{Introduction}

Wireless networks are subject to fundamental limitations in establishing source-destination data sessions. Investigating such limitations, along with discovering potential communication capabilities of wireless networks is of vital importance in designing efficient and practical algorithms for their operation. While Shannon's approach (\cite{Shannon} ) to mathematical analysis of communication systems is the most powerful approach, it is not easily extendable to wireless networks with large number of nodes. 

The pioneering work of Gupta and Kumar (\cite{Gupta} ) in 2000, ignited the efforts in characterizing the fundamental communication limits and capabilities of wireless networks. Gupta and Kumar's work, along with subsequent papers (\cite{ElGamal}, \cite{Kulkarni} and \cite{Fran} ), established the order of $\sqrt{n}$ achievable aggregate throughput for wireless networks with multihop technology, where $n$ is the number of nodes. However, in another line of research, the linear upper bound of order $n$ is derived for the capacity of wireless networks by exploiting information-theoretic max-flow min-cut discussions \cite{Jovicic}. The notable work of \"{O}zg\"{u}r et al. in 2007 resolves such gap between the upper and lower bounds \cite{Ozgur1}. In fact, by not considering interference as being always harmful, and by exploiting Multiple-Input Multiple-Output (MIMO) techniques, they propose a hierarchical cooperation scheme achieving linear throughput scaling.

Many of papers on wireless networks capacity, use channel models based on distance between nodes, while others use models based on random distributions. As also mentioned in \cite{Gowikar1}, in many scenarios, a wireless channel model based on randomness is a more appropriate choice than distance-based models. As an example of such scenario, we can point to the case where randomly moving obstacles block signal propagation, and the distance-based model cannot address such issues. Also, when the network area size is small, the dominating factor in characterizing the channel properties between nodes is the random fluctuations due to fading, rather than the distance-based path-loss effect. In addition, in such situation, the network is strongly interference-limited, which is best modeled by a random-based channel model. Moreover, many wireless systems employ a unit called Automatic Gain Control (AGC) which compensates for the distance effect. Accordingly, in many scenarios, it is more suitable to use a randomness-based channel model which is called the ``Random Connection Model''. In such model, the channel power $\gamma$ between each two nodes is drawn from a common parent distribution $f(\gamma)$, and different links are independent.

The first work considering the random connection model in communication over wireless networks is by Gowaikar et al. \cite{Gowikar1}. They propose a multihop scheme achieving linear scaling, for a specific case of parent distribution. Their scheme is based on establishing routes in random graphs. Their subsequent paper investigates a model which considers both the geometry and randomness effects \cite{Gowikar2}. Another work using the random connection model is the paper by Cui et al. \cite{Cui}. In their work, one-hop and two-hop communication schemes are investigated. It is shown that, in the class of parent distributions with finite mean and variance, the one-hop throughput is upper bounded by order $n^{1/3}$. Also, for two-hop schemes, they provide upper and lower bounds of order $n^{1/2}$.

While Cui et al. prove that in one-hop schemes, and in the class of parent distributions with finite mean and variance, one cannot surpass the throughput scaling of order $n^{1/3}$, they leave the achievability part unanswered. In this letter, we solve this open problem and propose an scheme achieving the throughput scaling of order $n^{1/3-\delta}$, for any $\delta>0$ and independent of $n$. Our proposed scheme is very simple and is based on establishing the largest number of concurrent communications.

The letter structure is as follows. In section II, the network model is explained. In section III, we explain the proposed scheme and prove that it achieves throughput of order $n^{1/3-\delta}$. Finally, section IV concludes the letter.

\section{Network Model}
Consider a wireless network consisting of $n$ nodes. Each node is capable of transmitting and receiving signals simultaneously (i.e., full duplex communication). The nodes follow an on/off strategy. In such strategy, at each time slot, a subset of nodes with $m$ elements are ``on'' and transmit simultaneously, while other nodes do not transmit any signal. We call the subset of active nodes $\mathbb{S}$. Each node in the network is a source of data for exactly one destination, and also, it is destination of data for exactly one source. Thus, we have $n$ sources (i.e., $S_1$, $\dots$, $S_n$), and $n$ destinations (i.e., $D_1$, $\dots$, $D_n$). Each source node $S_i$ wishes to transmit to the destination node $D_i$ for $i=1, \dots, n$. The signal received by $D_i$ at a specific time slot is:

\begin{equation}\label{eq1}
	y_i=\sum_{j\in \mathbb{S}}{h_{j,i}x_j}+n_i
\end{equation}
where $x_j$ is the signal transmitted by $j$th source node, and $h_{j,i}$ is the channel gain between $S_j$ and $D_i$. We define $\gamma_{j,i}\triangleq|h_{j,i}|^2$ to be the channel power, which is a random variable drawn from the parent distribution $f(\gamma)$. In addition, all links are independently and identically distributed (i.i.d.). Finally, $n_i$ is the additive white gaussian noise at each receiver whose variance is $N_0$.

The communication between $S_i$ and $D_i$ is successful, if and only if the received Signal to Interference and Noise Ratio (SINR) at $D_i$ is above a given threshold level:

\begin{equation}\label{eq2}
SINR_i \triangleq \frac {\gamma_{i,i}} {N_0+\sum_{j \in \mathbb{S}, j\neq i}{\gamma_{j,i}}} \geq \beta
\end{equation}

As explained earlier, such channel mode, also known as the ``Random Connection Model''  is a very appropriate model in many network scenarios \cite{Gowikar1}, \cite{Cui}.

\section{Throughput Achievability of Order $n^{1/3}$ }

We consider one-hop communication between sources and destinations. At each time slot, the nodes belonging to the active subset $\mathbb{S}$ broadcast their signals, and the rest of the nodes do not transmit. We define the one-hop throughput of the network as the expected number of successful receptions at each time slot (similar to \cite{Cui}). Cui et al. have proved that the throughput of such one-hop strategy, when $f(\gamma)$ has finite mean and variance, is upper bounded by order $n^{1/3}$. In this section, we propose an achievable scheme which achieves the throughput of order $n^{1/3-\delta}$ for any $\delta>0$ and independent of $n$. The main result of the letter is stated in the following theorem: \\

 \textbf{Theorem 1} \\
There exists a one-hop communication scheme achieving the throughput of order $n^{1/3-\epsilon/3}$ for any strictly positive $\epsilon$. The parent distribution resulting in this throughput is $f(\gamma)=\frac{2+\epsilon}{(1+\gamma)^{3+\epsilon}}$ for $\gamma\geq 0$, which has finite mean and variance.


 \textbf{Proof:} \\
Consider source nodes $S_1, \dots, S_n$ and destination nodes $D_1, \dots, D_n$. The channel power between $S_i$ and $D_i$ is $\gamma_{i,i}$. Let us sort the source and destination pairs based on the power of direct link between them (i.e., $\gamma_{i,i}$'s). Define $S_{(n-i+1)}-D_{(n-i+1)}$ as the source-destination pair which have $i$th most powerful channel, $\gamma_{(n-i+1),(n-i+1)}$. Thus, we have:

\begin{equation}\label{eq3}
	\gamma_{(1),(1)} \leq \gamma_{(2),(2)} \leq, \dots, \leq \gamma_{(n),(n)}
\end{equation}

In the proposed scheme, at each time slot, the first $m$ strongest source-destination pairs (i.e., $S_{(k)}-D_{(k)}, k=n-m+1, \dots, n$) are active, and other nodes are inactive. In other words, at each time slot, sources $S_{(k)}, k=n-m+1, \dots, n$ broadcast their signals simultaneously, and the corresponding receivers $D_{(k)}, k=n-m+1, \dots, n$ attempt to decode their messages. If we define $M$ as the number of successful receptions, by defining $r \triangleq n-m+1$, for the network throughput we have \footnote{$\mathbb{E}\{.\}$ and $\mathbb{P}\{.\}$ are the expectation operator and probability measure respectively.}:

\begin{eqnarray}\label{eq4}
\mathbb{E}\{M\}&=& \sum_{k=r}^{n}{\mathbb{P}\{ SINR_{(k)} \geq \beta \}} \\ \nonumber
&\geq& m\mathbb{P}\{SINR_{(r)} \geq \beta \} \\ \nonumber
&=& m \mathbb{P} \{\gamma_{(r),(r)}\geq\beta(N_0+\sum_{j=r+1}^{n}{\gamma_{(j),(r)}})  \} \\ \nonumber
&\geq& m \mathbb{P} \{\gamma_{(r),(r)} >2\beta\bar{\gamma}m\} \mathbb{P} \{ \beta(N_0+\sum_{j=r+1}^{n}{\gamma_{(j),(r)}}) < 2\beta\bar{\gamma}m \}
\end{eqnarray}

where $\bar{\gamma}\triangleq \mathbb{E}\{f(\gamma)\}$. The first inequality is due to the fact that $S_{(r)}-D_{(r)}$ has the weakest direct channel power among the active pairs. The last inequality is due to the independence of $\gamma_{(r),(r)}$ and $\beta(N_0+\sum_{j=r+1}^{n}{\gamma_{(j),(r)}})$. According to Markov's inequality we have:
\begin{eqnarray}\label{eq5}
	\mathbb{P} \{ \beta(N_0+\sum_{j=r+1}^{n}{\gamma_{(j),(r)}}) > 2\beta\bar{\gamma}m \} &\leq&  \frac{\beta(N_0+(m-1)\bar{\gamma})}{2\beta\bar{\gamma}m} \\ \nonumber 
&\simeq& \frac{1}{2} 
\end{eqnarray}
for large $m$. From (\ref{eq4}) and (\ref{eq5}) we have:
\begin{equation}\label{eq6}
	\mathbb{E}\{M\} \geq \frac{m}{2} \mathbb{P} \{\gamma_{(r),(r)} >2\beta\bar{\gamma}m\}
\end{equation}
At this stage of the proof, we need the following theorem due to Falk \cite{Falk}:

 \textbf{Theorem 2} \\
Suppose $X_1,\dots,X_n$ are $n$ i.i.d. random variables with the parent distribution $f(x)$. Define $X_{(1)},\dots,X_{(n)}$ to be the order statistics of these random variables. Suppose $F(x)$ is the cumulative distribution function (cdf) of the parent distribution, which is absolutely continuous, and for some $\alpha>0$ we have (von Mises condition \cite{Barry}):

\begin{equation}\label{eq7}
\lim_{x\rightarrow \infty} x\frac{f(x)}{1-F(x)}=\alpha
\end{equation}
\\
Then, if $ i \rightarrow \infty$ and $ i/n \rightarrow 0$ as $ n \rightarrow \infty$, there exist sequences $a_n$ and $b_n>0$ such that
\begin{equation}\label{eq10}
	\frac{X_{(n-i+1)}-a_n}{b_n} \Rightarrow N(0,1)
\end{equation}
where $\Rightarrow$ stands for convergence in distribution, and $N(0,1)$ is the normal distribution with zero mean and unit variance. Furthermore, one choice for $a_n$ and $b_n$ is:
\begin{eqnarray}\label{eq11}
	a_n=F^{-1}(1-\frac{i}{n}) \\ \nonumber
	b_n=\frac{\sqrt{i}}{nf(a_n)}
\end{eqnarray}

Now, we are ready to apply Theorem 2 to the throughput analysis of our scheme. In our scheme, we look for the statistical properties of $\gamma_{(r),(r)}$ to analyze $\mathbb{P} \{\gamma_{(r),(r)} >2\beta\bar{\gamma}m\}$, which appears in (\ref{eq6}) (note that $r=n-m+1$, and $m$ is the number of active sources.). Thus, we have the same ``intermediate order statistics'' problem as the one stated in Theorem 2. Consequently, we put:
\begin{equation}\label{eq12}
	X_{(n-i+1)}=\gamma_{(r),(r)}
\end{equation}
and
\begin{equation}\label{eq13}
	i=m=n^{\frac{1}{3}-\delta}
\end{equation}
for any $\delta>0$ and independent of $n$. Also, the probability distribution function (pdf) of the parent distribution which results in the desired throughput is:
\begin{equation}\label{eq14}
	f(x)=\frac{2+\epsilon}{(1+x)^{3+\epsilon}}, x\geq 0
\end{equation}
where $\epsilon>0$ is any small non-zero real number. This distribution has finite mean and variance.  Also, we observe that the corresponding cdf is absolutely continuous and satisfies the von Mises condition:
\begin{equation}\label{eq15}
\lim_{x\rightarrow \infty} x\frac{f(x)}{1-F(x)}=2+\epsilon >0
\end{equation}
Accordingly, due to theorem 2 we have:
\begin{equation}\label{eq16}
	\frac{\gamma_{(r),(r)}-a_n}{b_n} \Rightarrow N(0,1)
\end{equation}
where
\begin{eqnarray}\label{eq17}
	a_n&=&F^{-1}(1-\frac{i}{n}) \\ \nonumber
	&\simeq&n^{\frac{1}{3}\frac{2+3\delta}{2+\epsilon}} \\ \nonumber
	&=&n^{\frac{1}{3}}
\end{eqnarray}
where we have put $\delta=\epsilon/3$. Therefore, we have:
\begin{eqnarray}\label{eq18}
	\mathbb{P} \{\gamma_{(r),(r)} >2\beta\bar{\gamma}m\} &=& \mathbb{P} \{\gamma_{(r),(r)} >(2\beta\bar{\gamma}n^{-\delta})n^{1/3} \} \\ \nonumber
&>& \mathbb{P} \{\gamma_{(r),(r)} > n^{1/3}\} \\ \nonumber
&=&\frac{1}{2}
\end{eqnarray}
where the inequality is valid for large-enough $n$, due to the fact that $\bar{\gamma}$ and $\beta$ are independent of $n$. The last equality is a consequence of the result of Theorem 2, which is stated in equation (\ref{eq16}). By putting (\ref{eq18}) in (\ref{eq6}) we will have:
\begin{equation}\label{eq19}
	\mathbb{E} \{M\} \geq \frac{m}{4}
\end{equation}
where $m=n^{1/3-\delta}$, and Theorem 1 is proved.

$\hspace{3 in}$ $\Box$ 

\section{Conclusion}

In this letter, we have proved that the lower bound of one-hop communication in wireless networks with random connection model, in the class of finite mean and variance channel powers, is $n^{1/3-\delta}$, where $\delta>0$ is independent of $n$. Our result, combined withmises the upper bound of $n^{1/3}$ derived by Cui et al., characterizes the throughput capacity of such networks.

\section{Acknowledgement}
This work was supported in part by Iran National Science Foundation under Grant 87041174 and in part by Iran Telecommunications Research Center.

\bibliographystyle{ieeetr}

\end{document}